\documentclass[prl,letterpaper,twocolumn,amssymb,superscriptaddress,showpacs]{revtex4}
\usepackage{graphicx}

\newcommand{\mnras}{Mon.\ Not.\ R.\ Astron.\ Soc.\ }
\newcommand{\fs}{f_{\rm sat}}
\newcommand{\ms}{M_{\rm sat}}
\newcommand{\etal}{{\it et al.}}

\begin{document}

\title{Strong Lensing Constraints on Small-Scale Linear Power}

\author{Neal Dalal}
\affiliation{Physics Dept., UCSD 0350, 9500 Gilman Dr., La Jolla CA 92093}

\author{Christopher S. Kochanek}
\affiliation{Harvard-Smithsonian Center for Astrophysics, 60 Garden St.,
  Cambridge, MA 02138}

\begin{abstract}
We place limits on the linear power spectrum on small scales
($k\gtrsim 50h\hbox{ Mpc}^{-1}$) using measurements of
substructure in gravitational lens galaxies.  We find excellent
agreement with the simplest $\Lambda$CDM models, and in conjunction
with other cosmological probes, place constraints on the neutrino
mass $m_\nu$, tilt of the primordial power spectrum $n$, and mass of
the dark matter particle $m$. We find $n>0.94$, and for a
Harrison-Zeldovich spectrum, $m_\nu<0.74$ eV and $m>5.2$ keV, at 95\%
confidence. 

\end{abstract}


\maketitle

The amplitude of matter fluctuations on subgalactic scales can provide
a wealth of information on a wide range of physics.  For example,
the inflaton potential \cite{kl00}, the neutrino mass \cite{lyanu},
and the physics of the dark matter particles \cite{wdm,sidm} can all
cause measurable effects on the primordial power spectrum, especially 
on small scales.  Hence, measurement of the small scale linear power 
spectrum is of great interest.
Unfortunately, direct probes of the linear power on these
scales has proven difficult, due to a host of astrophysical processes
which complicate the interpretation of attempted measurements.  The
most promising avenue of attack has been study of the Lyman $\alpha$
forest absorption power spectrum \cite{lyagood}, however it has been
argued that inference of the linear power spectrum from the Ly
$\alpha$ forest will be difficult due to nonlinear effects
\cite{lyabad}.
One relatively clean probe of the linear power spectrum is
the number of collapsed halos as a function of mass.  Assuming
Gaussian density fluctuations, simple smoothing arguments may be
applied to calculate the collapsed halo abundance given the linear
power spectrum\cite{ps}.  This Press-Schechter formalism
reproduces the mass function produced in cosmological simulations
remarkably well, with only minor modifications \cite{st}.  A
consequence of Press-Schechter theory, subsequently confirmed by
numerical simulations, is that simple inflationary cold dark matter
(CDM) models predict halos with significant amounts of satellite
substructure \cite{substruc}.  The dearth of observed Local Group
dwarf satellites, relative to these predictions, could in principle
hint at suppression of fluctuations on small scales, implying
new physics in inflation \cite{kl00} or in the dark matter
sector \cite{sidm,wdm}.  Unfortunately, once again uncertainties in
the astrophysics of such dwarfs, such as star formation or feedback
processes \cite{bullocks}, limit the utility of these objects in
constraining the matter power spectrum.  Instead, a direct probe of
the satellite mass (as opposed to the light) is required.
Gravitational lensing, as an effect sensitive to mass rather than
light, can provide such a probe.  Recently, the abundance of satellite
substructure in a sample of elliptical galaxies has been measured
using gravitational lensing \cite{us}.  In this Letter, we quantify
the constraints of this detection on the linear power spectrum.

\begin{figure}
\includegraphics[width=0.45\textwidth]{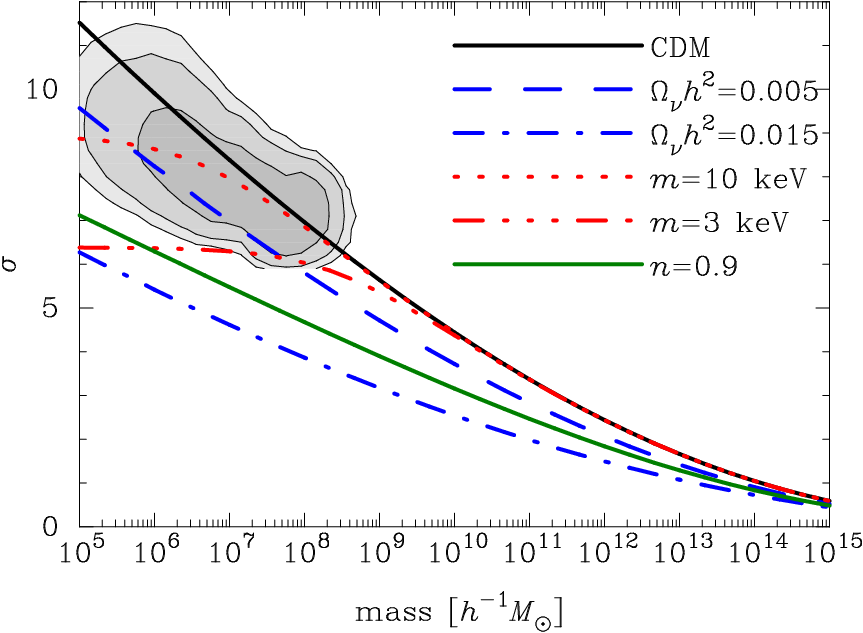}
\caption{Variance $\sigma$ on mass scale $M$.
Contours depict the 68\%, 90\% and 95\% confidence regions from
lensing.  The curves depict predictions for the cosmologies denoted
which, unless labeled otherwise, are flat $\Lambda$CDM universes with
$h=0.66$, $\Omega_{\rm M}h^2=0.15$, $\Omega_\nu=0$ and $n=1$. For the
WDM models, $m$ denots the WDM mass.
\label{sigma}}
\end{figure}

Gravitational lens galaxies require a significant fraction,
$0.006<\fs<0.07$, of their mass to be in subhalos 
in order to be consistent with observations of
lens image fluxes and relative positions \cite{us}.  
The most likely candidates for such satellites are the CDM subhalos
predicted by Press-Schechter arguments and seen in N-body simulations; see
Ref.~\cite{us} for more details.
Using the conditional mass function \cite{bower}, which has been shown
to match well the satellite mass function found in high resolution
N-body simulations \cite{kl00}, this $\fs$ directly
translates into a measurement of the linear rms density fluctuations on such
mass scales, $\sigma(\ms)$.  For a Sheth-Tormen (ST) multiplicity
function \cite{st}, we have  
\begin{equation}
\fs(>\ms)=A\left[{\rm erfc}\left(\sqrt{\frac{a\nu}{2}}\right)+
\frac{\Gamma(\case{1}{2}-p,a\nu/2)}{2^p\sqrt{\pi}}\right]
\end{equation}
\begin{equation}
\nu=\frac{\left[\delta_c\left(D(z_{\rm sat})^{-1}-D(z_{\rm
gal})^{-1}\right)\right]^2}{\sigma^2(\ms)-\sigma^2(M_{\rm gal})}.
\end{equation}
Here, $\delta_c=1.68$ is the critical linear overdensity for
collapse \cite{eke}, erfc($x$) is the complementary error function,
$\Gamma(n,x)$ is the incomplete gamma function,
$A^{-1}=1+2^{-p}\pi^{-1/2}\Gamma(\case{1}{2}-p)$ is a normalization
constant, $D(z)$ is the linear growth factor normalized so that
$D(z=0)=1$, and $z_{\rm gal}\approx1$ is a typical redshift at which
the lens galaxy halo of mass $M_{\rm gal}\approx 10^{12.5}M_\odot$
formed.  For a Press-Schechter multiplicity 
function, we would have $a=1$, $p=0$ and $A=\case{1}{2}$, however
ST find that N-body simulations are better fit by
$a=0.707$, $p=0.3$ and $A\approx0.322$ \cite{st}.  We estimate the
typical formation redshift for satellites by assuming they are tidally
truncated and setting their inferred density equal to 200 times the
background matter density when they formed.
Assuming a flat $\Lambda$CDM cosmology with $\Omega_M=0.33$ and Hubble
constant $h=H_0$/(100 km/s/Mpc)=0.66, and adopting the ST fit,
the lens constraints on satellite Einstein radius and mass fraction
translate into constraints on $\sigma(M)$.  This is plotted in
Figure~\ref{sigma}. 

Having determined the rms linear mass fluctuations on subgalactic
scales, we now compare this result to the predictions of various
cosmologies.  A representative sample are plotted in the figure.
Fortuitously, the wide degeneracy in the lens constraints happens to lie
nearly orthogonally to the direction in which different $\sigma(M)$
curves shift as the cosmology is varied.  
Clearly, the lensing measurement can strongly constrain a variety of
physics.  We consider three such constraints: the tilt of the
primordial power spectrum, the mass of the neutrino, and the mass of
the dark matter particle.  Throughout, we use COBE normalized power
spectra \cite{bunnwhite}, assume a flat $\Lambda$CDM cosmology with 
$\Omega_{\rm DM}h^2=0.13$, $\Omega_{\rm B}h^2=0.02$ and $h=0.66\pm0.05$, and
employ fitting functions for the transfer functions
\cite{eisensteinhu,wdm}.

The tilt of the primordial power spectrum, $n$, is related to the slow
roll parameters of the inflaton potential \cite{kolbturner}.  Recent
measurements \cite{cmb} of the cosmic microwave background radiation
constrain $n\approx 1\pm0.1\ (1~\sigma)$ 
on large scales ranging up to the horizon size.
Lensing probes mass scales some 18 orders of magnitude smaller, so it
provides the strongest constraints on the tilt that are currently
possible.  We find $0.93 < n < 1.04$ at 95.5\% confidence, consistent
with a scale-invariant Harrison-Zeldovich spectrum of perturbations.
If we instead derive one-sided limits, we find $n>0.94$ at 95.5\%
confidence.  
Inflationary models with broken scale invariance \cite{kl00,bsi} are
tightly constrained; the inflaton potential cannot have features which
break scale invariance on scales larger than $\sim 20$ kpc.

Experiments probing atmospheric \cite{superk} and solar neutrinos
\cite{sno} have
established that neutrinos have mass and thus constitute particle
(hot) dark matter.  However, these experiments do not provide a
definite measure of the mass of neutrinos, but instead the squared
mass difference $\Delta(m^2)$ between different neutrino flavors.  The
linear power spectrum constrains the neutrino mass, since
\cite{kolbturner,lyanu} 
\begin{equation}
m_\nu = 94\Omega_\nu h^2~\hbox{eV}.
\end{equation}
Note that this relation only applies for active neutrino flavors, and
for vanishing lepton asymmetry, however recent work indicates that the
asymmetry (and hence its corrections) are small \cite{dolgov}.  
Increasing the energy
density in neutrinos has the effect of suppressing power on small
scales.  Previous constraints \cite{lyanu}
from Lyman $\alpha$ forest observations limit $m_\nu<5.5$ eV.  By
probing even smaller scales, lensing further tightens this bound.  
We explored a range of neutrino masses ranging from 0.05 eV, near the
lower limit imposed by atmospheric neutrino oscillations \cite{superk}
up to 3.2 eV.  For a Harrison-Zeldovich spectrum,
and assuming a uniform prior on $m_\nu$, we find $m_\nu < 0.74$ eV at
95.5\% confidence.  If we instead assume a logarithmic prior, which
may be more appropriate, we find $m_\nu < 0.53$ eV at 95.5\% confidence.  
%
Interestingly, the lensing upper limit on the sum of neutrino masses
is quite close to the {\it lower} limit on neutrino mass measured by
the LSND experiment \cite{lsnd}, roughly $\sqrt{\Delta m^2}\sim0.3$ eV.

Additionally, our measurement provides a lower limit on the energy
spectrum of dark matter particles.  We follow standard convention and
characterize the energy spectrum by the mass of a neutrino with an
equivalent free streaming scale.  Warm dark matter models, with
$m\lesssim1$ keV, have been proposed as a modification to CDM, in
order to account for the dearth of dwarf satellite galaxies and the
lack of cusps in dark matter density in the central regions of
galaxies \cite{wdm}.  The close agreement of lensing observations with
CDM predictions indicates that the free streaming scale must be below
the $\sim20$ kpc scales probed by lensing.  Assuming a uniform prior
in the range 1 keV $<m<$ 32 keV, and assuming a Harrison-Zeldovich
spectrum, we find that $m>5.2$ keV is required at 95.5\% confidence.
Since the dark matter particle mass can be much larger than the range
we have explored, our bound is conservative.

\begin{figure*}
\centerline{
\includegraphics[width=0.3\textwidth]{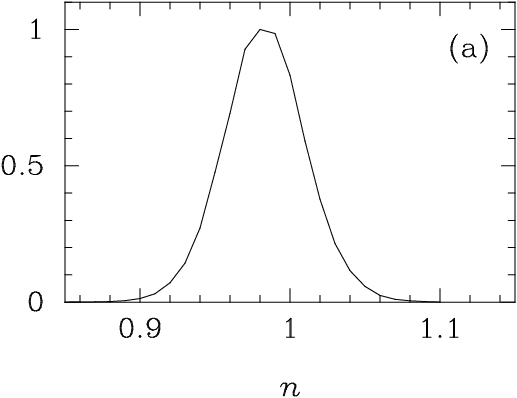}\hfil
\includegraphics[width=0.3\textwidth]{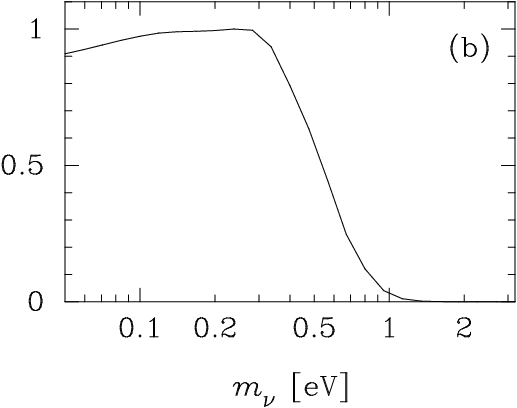}\hfil
\includegraphics[width=0.3\textwidth]{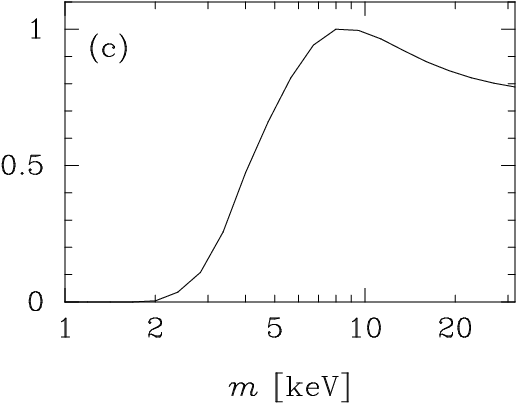}
}
\caption{Likelihood of (a) tilt of primordial power
spectrum, (b) neutrino mass, (c) dark matter particle mass.
\label{likely}}
\end{figure*}

Certain effects which we have not considered will tend to suppress the
effects of substructure, and hence bias our measurements to low
$\sigma$.  First, the conditional mass function ignores the effects of
tidal stripping and disruption of satellite subhalos.  Secondly, the
finite angular sizes of radio QSO's tend to wash out the lens
effects of halo substructure.  A proper treatment of these effects is
beyond the scope of this Letter, however we note that both effects
tend to increase the required abundance of satellite substructure and
strengthen the constraints.  Hence our limits should be regarded as
conservative. 

These constraints should all improve as the sample of suitable lens
systems expands, and as the quality of the data improves.  In 
general, multiply-imaged compact radio sources will be the 
preferred probe because of the high imaging resolutions 
available for radio sources using Very Long Baseline Interferometry
(VLBI).  For the current lenses, deep VLBI observations of 
extended, thin jet structures in lenses \cite{trotter00}
are a promising means of breaking the wide degeneracy between satellite 
mass and mass fraction.  The Extended Very Large Array \cite{evla}   
and the proposed Square Kilometer Array radio array \cite{ska} can
easily expand the sample of radio lenses by more than an order of 
magnitude, with a corresponding reduction in the uncertainties.

\bigskip

We thank Kev Abazajian, James Bullock, Joanne Cohn, Marc Davis, George
Fuller, Ravi Sheth, Max Tegmark and Martin White for helpful
discussions.  ND was supported by the Dept.\ of Energy under grant
DOE-FG03-97-ER 40546, and by the ARCS Foundation.  CSK was supported
by the Smithsonian Institution and NASA grants NAG5-8831 and
NAG5-9265.

\end{document}